# An Identity Based Strong Bi-Designated Verifier (t, n) Threshold Proxy Signature Scheme

*Sunder Lal and Vandani Verma*
*Department of Mathematics, Dr. B.R.A. (Agra), University,*
*Agra-282002 (UP), India*
E-mail- sunder_lal2@rediffmail.com, vandaniverma@rediffmail.com

**Abstract:** Proxy signature schemes have been invented to delegate signing rights. The paper proposes a new concept of Identify Based Strong Bi-Designated Verifier threshold proxy signature (ID-SBDVTPS) schemes. Such scheme enables an original signer to delegate the signature authority to a group of 'n' proxy signers with the condition that 't' or more proxy signers can cooperatively sign messages on behalf of the original signer and the signatures can only be verified by any two designated verifiers and that they cannot convince anyone else of this fact.

**Keywords:** ID Based Cryptography, Proxy Signatures, Threshold Proxy Signatures, Bilinear Pairing, Designated Verifiers.

## 1. Introduction

Certificate based cryptography allows a user to use an arbitrary string, unrelated to his identity, as his public key. When another user wants to use this public key, she has to obtain an authorized certificate that contains this public key. This creates the certificate management problem. To address this problem, Shamir [13] introduced the concept of ID based cryptography in 1984. In ID-based public key cryptography (ID-PKC) user's public key is derived from certain aspects of his identity (email address, phone no. etc.) and a trusted third party called key generating center (KGC) generates secret key for the users. Mambo et al [11] introduced the concept of proxy signatures in 1996. In a proxy signature scheme, an original signer delegates his signing capability to another user called proxy signer. Proxy signer signs message on behalf of the original signer, however proxy signatures are different from the original signatures. In the same year, Jakobsson et al [2] proposed the concept of designated verifier signatures (DVS). In DVS schemes, only the designated verifier can check the validity of the signatures but cannot convince any third party about the validity of the signatures. Saeednia et al [12] introduced the feature of strongness in DVS in 2003. Strong Designated Verifier Signature (SDVS) scheme forces the designated verifier to use his secret key at the time of verification. Since then several SDVS [5, 6, 10, 14] schemes have been proposed. In 2003, Desmedt [1] raised the problem of generating multi-designated verifier scheme. However, the first bi-designated verifier signature scheme using bilinear maps was proposed by Laguillaumie et at [9] in 2004. In 2006, the authors [7] proposed the ID-based strong bi-designated verifier signature scheme. They also proposed the first ID based strong bi-designated verifier proxy signature schemes in which the designated proxy signature can only be verified by the two designated verifiers using their secret keys. Zhang [15] and Kim et al [4] independently constructed a threshold proxy signature scheme. In a (t, n) threshold proxy signature scheme, the original signer delegates parts of his signing power to a group of n proxy signers such that t or more proxy signers pooling their shares of delegation can generate proxy signatures but any (t-1) or fewer proxy signers cannot create a valid proxy signature. The first ID based threshold proxy signature scheme was proposed by Xu et al [16] in 2004 and the first ID-based designated verifier threshold proxy signature scheme was proposed by Juan et al [3] in 2007. In such schemes, the designated verifier can only verify the threshold proxy signatures. The paper presents the extension of Juan et al [3] scheme to bi-designated verifier. In our proposed scheme, any of the two designated verifiers can check the validity of the threshold proxy signatures but they cannot convince any third party about the validity of the signature. Anyone of them can check the validity of the signatures even if he is not aware of other's identity. Our scheme is useful in the situations where the signature verifier does not want to rely on a single source for the trueness of the signatures.

The rest of the paper is organized as follows – section 2 contains some preliminaries about bilinear pairings and Gap Diffie Hellman group. In section 3 we present our ID-SBDVTPS scheme. In section 4 we analyze its security and concluding remarks in section 5.

## 2. Definitions
### 2.1 Bilinear pairings
Let $G_1$ be a cyclic additive group generated by P, whose order is a large prime number q and $G_2$ be a cyclic multiplicative group with the same order q. Let $e: G_1 \times G_1 \to G_2$ be a map with the following properties:

**Bilinearity:** $e(aP, bQ) = e(P, Q)^{ab}$ $\forall$ $P, Q \in G_1$ and $a, b \in Z_q^*$.
**Non-degeneracy:** $\exists$ $P, Q \in G_1$, such that $e(P, Q) \neq 1$, the identity of $G_2$.
**Computability:** There is an efficient algorithm to compute $e(P, Q)$ $\forall$ $P, Q \in G_1$.
Such pairings may be obtained by suitable modification in the Weil-pairing or the Tate-pairing on an elliptic curve defined over a finite field.

### 2.2 Computational problems
**Decisional Diffie-Hellman Problem (DDHP):** Given $P, aP, bP, cP$ in $G_1$, decide whether $c = ab \mod q$.
**Computational Diffie-Hellman Problem (CDHP):** Given $P, aP, bP$ in $G_1$ compute $abP$
**Bilinear Diffie-Hellman Problem (BDHP):** Given $P, aP, bP, cP$ in $G_1$ compute $e(P, P)^{abc}$ in $G_2$.
**Gap Diffie-Hellman Problem (GDHP):** A class of problems, where DDHP can be solved in polynomial time but no probabilistic algorithm exists that can solve CDHP in polynomial time.

## 3. Identity Based Strong Bi-Designated Verifier (t, n) Threshold Proxy Signature Scheme.

Our scheme is an extension of Juan et al [3] scheme. The single designated verifier is extended to bi-designated verifier to form our ID-SBDVTPS scheme. In our scheme, we have assumed Alice as the original signer, PS = {$P_1, P_2,…P_n$} as the group of 'n' proxy signers and Bob and Cindy as the two designated verifiers and KGC stands for key generating centre. The scheme is divided into six stages: *setup, key-generation, secret-share generation, proxy-share generation, proxy-signature generation and proxy signature verification.*

- ❖ **Setup:** For a given security parameter k, $G_1$ is a GDH group prime order $q > 2^k$ generated by P and e: $G_1 \times G_1 \to G_2$ is a bilinear map. KGC chooses a master key $s \in Z_q^*$ and sets $P_{pub} = sP$. Chooses two cryptographic hash functions $H_1: \{0,1\}^* \to Z_q^*$, $H_2: \{0,1\}^* \times G_1 \to Z_q^*$ and $H_3: \{0,1\}^* \times G_1 \times G_2 \to Z_q^*$. The system parameters $(q, G_1, G_2, e, P, P_{pub}, H_1, H_2, H_3)$ are made public and 's' is kept secret with KGC.
- ❖ **Key generation:** Given a users identity ID, KGC computes his public key $Q_{ID} = H_1(ID)$ and the associated secret key $S_{ID} = s^{-1}Q_{ID}.P$.

- **Secret share generation:** The proxy group applies a (t, n) verifiable secret sharing scheme to generate secret shares for all the proxy signers in PS as follows:
  - Each $P_i \in PS = \{P_1, P_2, ...P_n\}$ randomly chooses a *(t - 1)* degree polynomial $f_i(x) = \sum_{l=1}^{t-1} a_{il} x^l + a_{io}$ with random coefficients $a_{il} \in Z_q^*$ and publishes $A_{il} = a_{il} P$, $l = 0, 1, 2, ...t – 1$. $P_i$ sends $f_i(j)$ to $P_j$ via a secure channel for $j \ne i$.
  - On receiving $f_i(j)$, $P_j$ can validate it by checking the equality
    $f_i(j)P = \sum_{k=0}^{t-1} j^k A_{ik}$, If it holds, each $P_i$ computes his secret share $r_i = \sum_{k=1}^{n} f_k(i)$ and publishes $U_i = r_i P$.

- **Proxy share generation:** Each proxy signer $P_i \in PS$ gets his own proxy signing key share as follows:
  - The original signer Alice first randomly chooses $r_w \in Z_q^*$ and computes
    $U_w = r_w Q_{IDA} P$, $h_w = H_2(m_w, U_w)$, $V_w = (r_w + h_w) S_{IDA}$
    The signature on $m_w$ is $w = (U_w, V_w)$. Finally, Alice sends $w$ and $m_w$ to each $P_i \in PS$
  - To verify a signature, the proxy signer $P_i$ computes $h_w = H_2(m_w, U_w)$ and accepts the signature iff $e(P_{pub}, V_w) = e(P, U_w + h_w Q_{IDA} P)$ and rejects it otherwise. If the signature $w$ is accepted, $P_i$ computes $S_i = S_{IDi} + V_w$ as his own proxy secret.
  - $P_i$ randomly chooses a *(t - 1)* degree polynomial $g_i(x) = \sum_{l=1}^{t-1} b_{il} x^l + S_i$ with random coefficients $b_{il} \in G_1$ and publishes $B_{il} = e(P, b_{il})$ for $l = 1, 2, ...t-1$. $B_{io}$ can be computed by each proxy signer as $B_{io} = e(P, U_w + (Q_{IDPi} + h_w Q_{IDA})P)$. Furthermore, $P_i$ sends $g_i(j)$ to $P_j$ via a secure channel for $i \ne j$.
  - On receiving $g_j(i)$, $P_i$ can validate it by checking the equality $e(P_{pub}, g_j(i)) = \prod_{k=0}^{t-1} B_{jk}^{i^k}$
    Finally, $P_i$ computes his proxy signing key share $SK_{Pi} = \sum_{k=0}^{t-1} g_k(i)$ and publishes $e(P_{pub}, SK_{Pi})$.

- **Proxy signature generation:** Let $D = \{P_1, P_2, ...P_t\}$ be the group of 't' proxy signers who want to sign message 'm' on behalf of the original signer Alice.
  - Apply the Lagrange interpolation formula to compute
    $X = Q_{IDB} Q_{IDC}$, $G_{Vi} = e(XP, S_{IDi})$, $Y_i = G_{Vi}^{r_i}$, $Y = \prod_{i=0}^{t} Y_i^{\eta_i}$, $\eta_i = \prod_{j \ne i}^{j \in \{1,2,...t\}} \frac{j}{j-i}$, $U = \sum_{i=1}^{t} \eta_i U_i$
    Let $H = H_3(m, U, Y)$. Each $P_i \in D$ computes $V_i = U_i + H\, SK_{Pi}$ and $\sigma_i = (U_i, V_i)$ be his own proxy signature share.
  - On receiving $\sigma_i$, the designated clerk validates it by checking $e(P, V_i) = e(P, U_i)\, e(P, SK_{Pi})^H$ If it holds, then $\sigma_i$ is the valid individual proxy signature share on *'m'*. If all the individual proxy signature shares for 'm' are valid, then the clerk computes $V = \sum_{i=1}^{t} \eta_i V_i$. The proxy signature on 'm' is σ = *(m, V<sub>w</sub>, m<sub>w</sub>, U, V)*

- ❖ **Proxy signature verification:** To verify the proxy signature σ, the designated verifiers Bob (and Cindy) compute $Q_{IDC} = Q_{IDB}^{-1} X$, (Bob) $Y^* = e(S_{IDB}Q_{IDC}, U(\sum Q_{IDPi}))$ and accepts the signature iff $e(P_{pub}, V) = e(P_{pub}, U + nHV_w) e(P, (\sum Q_{IDPi})P)^H$.

## 4. Security analysis:
In this section we analyze the security of the proposed ID-SBDVPS schemes.

**4.1 Correctness:** The following equation gives the correctness of the scheme for Bob

$$e(P_{pub}, V) = e(P_{pub}, \sum_{i=1}^{t} \eta_i V_i)$$

$$= e(P_{pub}, \sum_{i=1}^{t} \eta_i (U_i + H.SK_{Pi}))$$

$$= e(P_{pub}, \sum_{i=1}^{t} \eta_i U_i) e(P_{pub}, \sum_{i=1}^{t} \eta_i SK_{Pi})^H$$

$$= e(P_{pub}, U) e(P_{pub}, \sum_{i=1}^{t} \eta_i \sum_{k=1}^{n} g_k(i))^H$$

$$= e(P_{pub}, U) e(P_{pub}, (\sum S_{IDPi} + nV_w).H)$$

$$= e(P_{pub}, U) e(sP, s^{-1} \sum Q_{IDPi} P)^H e(P_{pub}, V_w)^{nH}$$

$$= e(P_{pub}, U) e(P, P)^{(\sum Q_{IDPi})H} e(P_{pub}, V_w)^{nH}$$

$e(P_{pub}, V) = e(P_{pub}, U + nHV_w) e(P, (\sum Q_{IDPi})P)^H$

**4.2 Strongness:** In the proposed scheme proxy signatures are generated in such a manner that only the two designated verifier Bob and Cindy can check the validity of the signatures using his secret key. Hence, our scheme provides the strongness property.

**4.3 Proxy protected:** Alice cannot generate a valid signature share on behalf of $P_i$, since he does not have any information about the secret key $S_{IDPi}$ of each $P_i$. Hence, our scheme is proxy protected.

**4.4 Secrecy:** In our proposed scheme, the original signer Alice secret key cannot be derived from any information such as the shares of the proxy signing key, proxy signature etc. Even if 't' out of 'n' proxy signers collaborates to deliver the proxy share, they cannot calculate the Alice secret key. Hence, our scheme is secure.

## 5. Conclusion:

In this paper, we have presented a new concept of Identity based strong bi-designated verifier (t, n) threshold proxy signature scheme. The proposed scheme can also be viewed as a double threshold signature scheme as it uses threshold in signature generation and signature verification phase. The scheme is applicable in the situations where receiver wants the signatures to be verified by two designated persons and no one other than these two designated persons can check the trueness of the signatures.